# A NLTE analysis of the hot subdwarf O star BD+28°4211.
# II. The optical spectrum

M. Latour[1], G. Fontaine[2], E.M. Green[3], and P. Brassard[2]

[1] Dr. Karl Remeis-Observatory & ECAP, Astronomical Institute, Friedrich-Alexander University Erlangen-Nuremberg, Sternwartstr. 7, 96049 Bamberg, Germany

[2] Département de Physique, Université de Montréal, Succ. Centre-Ville, C.P. 6128, Montréal, QC H3C 3J7, Canada

[3] Steward Observatory, University of Arizona, 933 North Cherry Avenue, Tucson, AZ 85721, USA

## ABSTRACT

We present the second part of our detailed analysis of the hot sdO and spectroscopic standard star BD+28°4211 where we focus on the optical spectrum. This target was selected in order to revisit the more general question of how reliably the atmospheric parameters of a very hot star can be inferred from optical spectroscopy alone. Given its status as a spectrophotometric standard, spectral data of exceptional quality are available for BD+28°4211. In the first part of our study, using UV and FUV spectra of the star, we determined the abundances of some 11 metals detected in the atmosphere of BD+28°4211 and corroborated the fundamental parameters estimated in past studies ($T_{\text{eff}} \sim 82,000$ K, $\log g \sim 6.2$, and solar $N(\text{He})/N(\text{H})$). In this work, we aim at rederiving those secured parameters on the sole basis of high quality optical spectra. A first grid of NLTE line-blanketed model atmospheres, including C, N, O, Mg, Si, S, Fe, and Ni with the abundances derived from the UV spectrum, does not give satisfactory results when we apply a standard simultaneous fitting procedure to the observed H and He lines of our optical spectra. The line profiles are not finely reproduced and the resulting effective temperatures, in particular, are too low by ~10,000 K. We next investigate the probable cause of this failure, i.e., the importance of missing opacity sources on the atmospheric stratification. We thus compare line profiles computed from models with artificially-boosted metallicities, from solar abundances to 15× those values. We find that the structural effects saturate for a metallicity of ~10× solar, and use this to compute a second full grid of models and synthetic spectra. This metal-enriched grid allows us to achieve significantly improved spectral fits with models having the same parameters as the expected ones within their uncertainties. As an a posteriori test, we compared the detailed profiles of several model spectral lines with high resolution spectra culled from archived HIRES observations. The agreement between our synthetic lines and the observed ones is very convincing. Our test case thus reveals that there is still a need for models with enhanced metallicity for better estimating the atmospheric parameters of objects such as hot subdwarfs and hot white dwarfs if only optical spectra are available.

**Key words.** stars : atmospheres — stars : fundamental parameters — stars : individual (BD+28°4211) — subdwarfs

## 1. Astrophysical Context

Hot subluminous O stars form the hottest part of the extreme horizontal branch (EHB) region, which is itself a hot extension of the horizontal branch. The EHB region of the Hertzsprung-Russell (HR) diagram encompasses stars that span a wide range of effective temperatures, from 22,000 K up to 100,000 K, and are more compact ($4.8 \lesssim \log g \lesssim 6.4$) than the main sequence ones. This includes stars from two distinct spectral types, the cooler sdBs with their strong Balmer lines (22,000 K $\lesssim T_{\text{eff}} \lesssim$ 38,000 K), and the hotter sdOs showing strong He II lines as well ($T_{\text{eff}} \geq 38,000$ K). Most hot subdwarfs are believed to be helium core burning objects – or in a phase following immediately core helium exhaustion – with a layer of H-rich material too thin to sustain significant shell burning[1]. From a spectroscopic point of view, sdB stars form a rather homogeneous group: they mostly cluster within the theoretical core helium burning region and its immediate surroundings in the HR diagram. Numerous analyses were made on various samples of sdB stars and their global properties (helium content, metal content, rotational velocity, bi-nary population, etc.) are now well documented (e.g., Edelmann et al. 2003; Geier et al. 2010; Geier & Heber 2012; Geier 2013; Fontaine et al. 2014). However the situation is different when speaking of the hotter sdOs; they are distributed among a much larger region in the HR diagram and this distribution is not as homogeneous as in the case of the sdBs. While most of the latter are helium poor, the majority of sdO stars has an atmosphere enriched in helium and they are thought to be the results of various peculiar evolutionary paths[2].

One striking fact about sdO stars is that they have been much less studied than their coolest counterparts. The most significant study in our view has been the one carried out by Stroeer et al. (2007) using an homogeneous sample of sdOs observed within the SPY survey. Additional abundances of carbon and nitrogen were then measured in those sdO stars by Hirsch (2009) and Hirsch & Heber (in preparation, 2015). The sample of ω Centauri EHB stars in Moehler et al. (2011) and Latour et al. (2014c) also included a fair amount of sdO stars, but mostly cooler ones found at the transition between the spectral types B and O (i.e., below 40,000 K). While known sdO stars in the field are outnumbered by sdBs (these stars show a number ratio ≈ 1 :

---

[1] Note that a few hot subdwarfs have been found to have a mass too low to sustain helium core burning (e.g., HD 188112, Heber et al. 2003). However, these low mass stars represent only a tiny fraction of the hot subdwarf population.

[2] A comprehensive review of the global properties and characteristics of hot subdwarf stars can be found in Heber (2009).

3; Heber 2009), the true reason for this relative lack of investigations must be found in the inherent challenge associated with the atmospheric modeling and spectroscopic analyses of very hot stars. For a star having $T_{eff} \gtrsim 50,000$ K, the fundamental parameters determined by comparing the observed Balmer and helium lines in the optical with model ones bear important uncertainties. This is mostly due to the so-called Balmer line problem, first noticed by Napiwotzki (1992, 1993) in hot central stars of old planetary nebulae. Basically, this problem comes down to the inability to simultaneously reproduce the observed Balmer lines with a unique set of fundamental parameters (log $g$ - $T_{eff}$). More specifically, the individual lines need different temperatures in order to be matched properly, with the higher lines in the series needing models at higher temperatures. For example, for BD+28°4211, H$\alpha$ was best reproduced at $T_{eff} \simeq 50,000$ K and H$\epsilon$ at around 85,000 K (Napiwotzki 1993). In a situation like that, it is rather tricky to determine the temperature of the star without any additional information. In this particular case, the author could rely on UV data (from International Ultraviolet Explorer IUE) whose first analysis led to a value of $T_{eff} \approx 82,000$ K (Dreizler & Werner 1993). This relatively high value of $T_{eff}$ was also supported by the weakness of the H I 5876 Å line in the optical domain, which requires a high effective temperature. On the basis of these results, it was then concluded that the H$\epsilon$ line was the one that could provide the most realistic temperature estimate.

This "calibration" may have been useful at times, but it was not at all satisfactory on general grounds and different hypotheses were soon investigated to solve this embarrassing Balmer line problem (Napiwotzki & Rauch 1994). Most of them were rapidly rejected, save for the idea that the inclusion of metallic elements in the models might influence in a significant way the atmospheric structure, which in turn could change the shapes of the Balmer lines[3]. Note that at the time this issue was first identified, model atmospheres used for these hot stars were using a non-LTE (NLTE) treatment but included only H and He; the treatment of line blanketing by metals was still in its early stages. Accounting for the effects of metallic elements via their colossal numbers of transition lines was, at the time, a real computational challenge. Thanks to the work of Dreizler & Werner (1993) and Hubeny & Lanz (1995) on the development of numerical techniques allowing the inclusion of metals (such as C, N, O, and iron-group elements) and the treatment of their transition lines in NLTE calculations, it was subsequently shown that these elements can indeed influence strongly the thermodynamical structure of the atmosphere. However, the resulting effects on the Balmer lines were initially found to be surprisingly weak (Haas et al. 1996). It is Werner (1996) who brought up an important refinement in the treatment of light metals opacity: the inclusion of Stark broadening profiles for the CNO elements instead of the Doppler ones previously used. This addition led to an improved reproduction of the Balmer lines in his two test stars, BD+28°4211 itself and LS V+46°21, the DAO-type central star of a planetary nebula. Despite this breakthrough, hot stars such as those presented in Werner (1996) were then never analyzed by attempting a simultaneous fit of all of the available Balmer and helium lines in optical spectra. This now widely used technique has proven itself to be a robust tool for the determination of fundamental parameters ($T_{eff}$, log $g$, and sometimes also $N(He)/N(H)$) in cooler white dwarfs and sdB stars (Bergeron et al. 1994; Saffer et al. 1994). Rauch et al. (2007) later carried out a comprehensive spectral analysis of LS V+46°21, but they determined the effective temperature of the star using mainly the ionization equilibria of different metallic species whose lines were visible in the UV spectra of the star. The strongest He II lines ($\lambda\lambda$1640, 4686) and H$\beta$ were used to constrain the surface gravity. Feige 110 and G191-B2B were also analyzed in a similar way, combining both UV and optical data to assess fundamental parameters and metal abundances (Rauch et al. 2013, 2014). As for BD+28°4211, no further detailed studies were made on that star since Haas et al. (1996) and later on Ramspeck et al. (2003) estimated the abundances of a few metallic elements using IUE and HST Space Telescope Imaging Spectrograph (STIS) data.

Given the particular status of BD+28°4211 as a spectroscopic standard star, both in the optical domain as well as in the UV range, modern data of extremely good quality are publicly available (through the Mikulski Archive for Space Telescopes, MAST[4]). Surprisingly, and until recently, these data have merely been exploited. Moreover, some X-ray emission has been measured in BD+28°4211, as well as in two other sdO stars, Feige 34 and BD+37°1977 (La Palombara et al. 2014). In view of this state of affair, and given the availability of optical spectra of exceptionally high sensitivity (see below), we undertook an in-depth spectral analysis of this star with the main aim of testing the simultaneous optical fitting method in a very hot star. This is of importance for hot stars, the majority of them in fact, for which only optical spectroscopy is readily accessible while no UV data are available.

The first part of this analysis (Latour et al. 2013, hereafter Paper I) focussed on the UV spectral distribution of BD+28°4211 using STIS and Far-Ultraviolet Spectroscopic Explorer (FUSE) spectra. We obtained in a self-consistent way the abundances of 11 elements with well defined lines in the UV, namely C, N, O, F, Mg, Si, P, S, Ar, Fe, and Ni. None of these elements was found to be enriched, the abundances rather lie between the solar value and 1/10 solar. With the help of the ionization equilibria of several metallic species, we were able to confirm the previously determined effective temperature and constrain it to a value of $82,000 \pm 5,000$ K. We also estimated conservatively the surface gravity of the star to be log $g$ = $6.2^{+0.3}_{-0.1}$, which is also consistent with past results. By comparing the Hipparcos parallax measurement of BD+28°4211 (van Leeuwen 2007) with spectroscopic distances estimated from several model spectra we found that, in order to reconcile both values, the star needs either a log $g$ higher than 6.2 or a mass significantly lower than the canonical value of 0.5 $M_\odot$.

Having these informations at hand, we can now tackle the analysis of its optical spectrum. Past spectroscopic studies of hot stars like BD+28°4211 always relied on UV data, sometimes supported by optical ones, to get reliable fundamental parameters (e.g., Rauch et al. 2007; Fontaine et al. 2008; Ziegler et al. 2012). However, the need to rely on UV data can be very restricting since they must be gathered with space missions, which are a lot less accessible than ground-based observations supplying optical spectra. Our goal here is to find a way, using our test case star, to obtain reliable fundamental parameters ($T_{eff}$, log $g$ and $N(He)/N(H)$) using solely optical data. Given that changes in optical line profiles may be subtle at times in the very hot star regime, this necessitates spectra of high S/N ratio and/or high resolution. In this spirit, we exploit three very high sensitivity spectra having various resolutions and wavelength coverages. We exploit as well a high resolution UVES spectrum culled from the ESO archive. This material is described in more detail in the following section. The main part of this paper, Section 3, includes a description of the model grids as well as the subsequent spectroscopic analyses made. We also carried out some additional verifications to test our deduced fundamental parameters by comparing our best-fit models with additional high resolution archive spectra. Finally, a discussion follows in Section 5.

---

[3]  The suggestion that metal opacity is at the heart of the Balmer line problem was first made by Bergeron et al. (1993).

[4]  http://archive.stsci.edu/

## 2. Observational Material

BD+28°4211 is a well known, bright (V = 10.58) standard star and, as such, has been regularly observed for calibration purposes. In particular, as part of her spectroscopic programs at the University of Arizona, one of us (E.M.G.) has observed that standard star for many years using mainly three different instrumental setups, each corresponding to a different spectral resolution and spectral coverage. Hence, by carefully combining the individual calibration data for each setup, we have obtained three exceptionally high sensitivity spectra for BD+28°4211 on which is based a large part of the present analysis. This issue of the signal-to-noise ratio (S/N) is quite important in the present context since we seek to detect differences between the observed and modelled line profiles that may be relatively small.

It should also be pointed out that particular care has always been taken while observing BD+28°4211 in order to avoid contamination from the light of a nearby star. Indeed, rotating the slit to the parallactic angle at the midpoint of the exposure ensures that no light from the faint red companion of the star (Massey & Gronwall 1990) contaminates the spectrum of the sdO. Usually the companion was off the slit, however in the few cases when it fell within the slit, there was a clear spatial separation between the spectra of the companion and BD+28°4211, so it was always possible to extract only the sdO spectrum.

Our first instrumental setup is defined by the combination of the blue spectrograph attached to the 6.5 m Multiple Mirror Telescope (MMT). The 832 mm$^{-1}$ grating is used in second order and, with the choice of a 1″ slit width, this combination provides a resolution $R$ of ~4250 (1.0 Å) and covers the wavelength range 4000–4950 Å. The careful combination of 20 individual spectra of BD+28°4211 observed at the MMT resulted in a spectrum having a total S/N around 1,100[5]. This will be referred to as the MMT spectrum in what follows.

The other two instrumental setups make use of the Boller & Chivens (B&C) Cassegrain spectrograph mounted to Steward Observatory's 2.3 m Bok Telescope at Kitt Peak. Hence, the 832 mm$^{-1}$ grating in second order with a 1.5″ slit is used to achieve a resolution of 1.3 Å over a bluer wavelength range of 3675–4520 Å. Twenty spectra were obtained with this particular setup, each flux calibrated and then combined with median filtering. The resulting S/N ratio is ~ 918. This spectrum will later be referred to as the BOK1.3 one.

The third set of observations covers a much wider wavelength interval, from 3620 to 6900 Å, but at the cost of a lower resolution of 8.7 Å. These observations are still very useful because they include two additional and important spectral lines in BD+28°4211 : He II at 5412 Å and H$\alpha$. These low resolution spectra are obtained with a 400 mm$^{-1}$ grating in first order in conjunction with a 2.5″ slit. Our resulting 8.7 Å spectrum is the combination of 90 individual observations and has an overall whopping S/N of ~2500. The resulting spectrum is referred to as the BOK8.7 one in what follows.

Additionally, we retrieved one UVES spectrum of BD+28°4211 through the ESO archive[6] (program ID 69.C-0171(A)), that provides access to reduced scientific data obtained with the UVES spectrograph mounted on the VLT. That spectrum has a much reduced S/N ratio than our Steward Observatory (SO) data but it has significantly better resolution. It comes in two parts, 1) a "blue" one characterized by a spectral coverage of 3281–4562 Å, a resolving power R = 68,642, leading to $\Delta\lambda$ ~0.06 Å in mid-range, and a value of S/N ~ 95, and 2) a "red" one characterized by a spectral coverage of 4624–6686 Å, a resolving power R = 107,200, leading to $\Delta\lambda$ ~0.05 Å in mid-range, and a value of S/N ~ 63. One other important property of this UVES spectrum is that its continuum behaves relatively well for an echelle spectrum, so the observed line profiles are already amenable to direct comparisons with model line profiles. Thus, we made no particular attempt to remove some of the small remaining wavy structure shown in the data, except for fixing a discontinuity that is present in the red wing of the He II 4686 Å line. We refer to that spectrum as UVES in the rest of this paper.

There are also many high resolution HIRES observations of BD+28°4211 available in the Keck Observatory Archive[7] and extracted spectra, produced by an automated pipeline, are available for more than a hundred of them. Due to continuum placement difficulties, however, these echelle data are not particularly suited for a formal analysis aimed at simultaneously fitting the optical lines of hydrogen and helium. Nevertheless, some of the HIRES spectra are of very good quality (S/N up to ~300) and are highly interesting since they feature their fair share of details that cannot be seen in our own spectra. We thus retrieved 23 of the available extracted spectra ($\lambda$ between 4600 - 6600 Å, highest S/N) in order to make a posteriori comparisons between some observed lines and our optimal model spectra as well as to look for any radial velocity variations. Among those spectra, two were featured in Herbig (1999). However, there is a major inconvenience with the HIRES spectra: the continuum of the various orders is uneven and quite a bit wavy. While this flaw can be overcome rather easily when studying narrow spectroscopic features, which is an important purpose of such high resolution observations (~ 0.1 Å), it is a hard thing to deal with when the lines of interest are tens of angstroms in width. The best way we found to flatten the continuum of the retained spectra is by using the continuum of adjacent orders which often had a similar shape. Continua from adjacent orders were shifted and superimposed to the spectra of interest and if both corresponded well enough, dividing the spectra by the continuum would flatten the former in a satisfactory way.

## 3. Spectroscopic Analysis

### 3.1. Model Atmosphere Grids

Our grids of models were computed with the public codes TLUSTY and SYNSPEC[8] (Lanz & Hubeny 1995), which were run on CALYS, our cluster of computers currently containing 320 processors, where a large number of models can be simultaneously computed. Further technical details on the models can

---

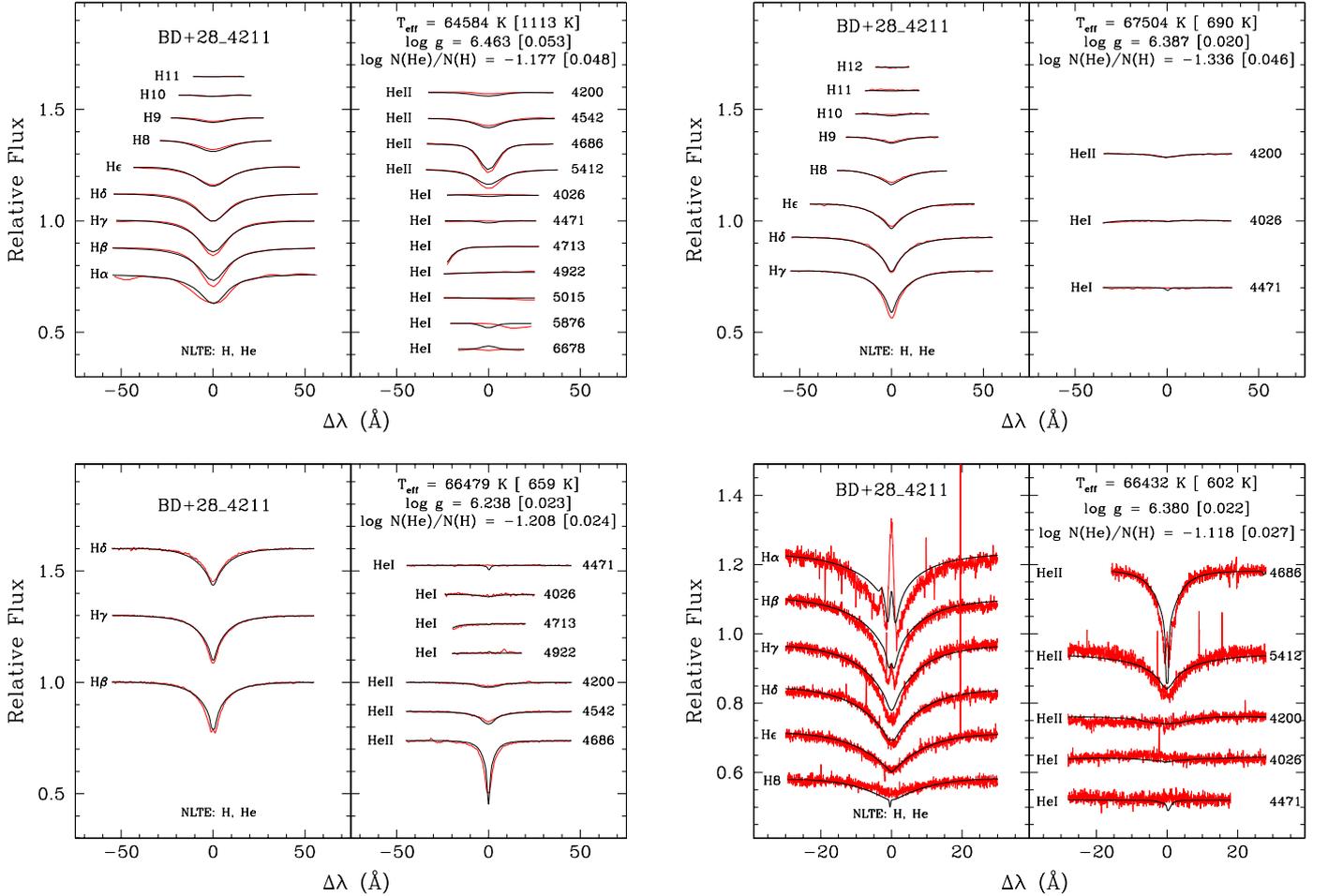

**Fig. 1.** Best fits obtained using our grid of NLTE metal-free models. In order of increasing resolution: BOK8.7 spectrum (top left), BOK1.3 spectrum (top right), MMT spectrum (bottom left), UVES spectrum (bottom right). The observed spectral lines are shown in red, while the modeled lines are shown in black.

be found in Paper I, especially about the ionic species that were included. We started our analysis with two different model grids. The first one is a metal-free grid of NLTE models; one of the grids that were built at the time of the analysis of the pulsating sdO star SDSS J160043.6+074802.9 (Latour et al. 2011). The purpose of using this grid is mainly for comparison. The second grid is one especially suited for BD+28°4211, that includes eight of the main metallic constituents of the star's atmosphere, namely C, N, O, Mg, Si, S, Fe, and Ni. Let us remind the reader here that in the course of the analysis made in Paper I, we inspected our model atoms and added a classic Stark profile to a few important lines of C IV, N IV, O IV-V and Si IV. The abundances of these elements were taken from the results of the UV analysis made in Paper I. Moreover, the hydrogen broadening profiles of Tremblay & Bergeron (2009) were added to the SYNSPEC code, as a replacement for the Lemke (1997) broadening tables previously used. This grid covers a parameter space centered around those of BD+28°4211, with $T_{eff}$ varying from 76 kK up to 90 kK by steps of 2,000 K, log $g$ from 5.4 to 6.8 dex by steps of 0.2 dex, and finally log $N$(He)$/N$(H) from −2.0 to 0.0 dex by steps of 0.5 dex.

We did not include F, P, and Ar in the metallicity considered for the construction of this grid mainly for technical reasons. Indeed, our current implementation of TLUSTY on our cluster CALYS leads to some convergence problems as well as memory restrictions when multi-metal NLTE model atoms are simultaneously considered. This is particularly true when complex atoms such as those of Fe and Ni are included (as is the case here). This considerably slows down the computations to the point of being impractical. Extensive tests have shown that, for the abundances deduced in Paper I, the three above elements contribute, in fact, negligibly to the overall metal opacity in the atmosphere of BD+28°4211 compared to the other elements that we retained. In this way, we were able to built our second grid over a reasonable length of time. Still, this second grid of models must be viewed as one that includes a *minimal* metallicity since it does not include all the metallic species present in the atmosphere of the star.

## 3.2. Derived Atmospheric Parameters

The four optical spectra of BD+28°4211 that we gathered were analyzed with the two model grids mentioned in the previous section. We stress that the spectra were analyzed in the same way as are usually handled the much cooler sdB stars: all of the available H and He lines being simultaneously fitted in a three-dimensional space ($T_{eff}$, log $g$, and log $N$(He)$/N$(H)). The $\chi^2$ minimization procedure relies on the method of Levenberg-Marquardt, based on a steepest descent method (Bergeron et al.

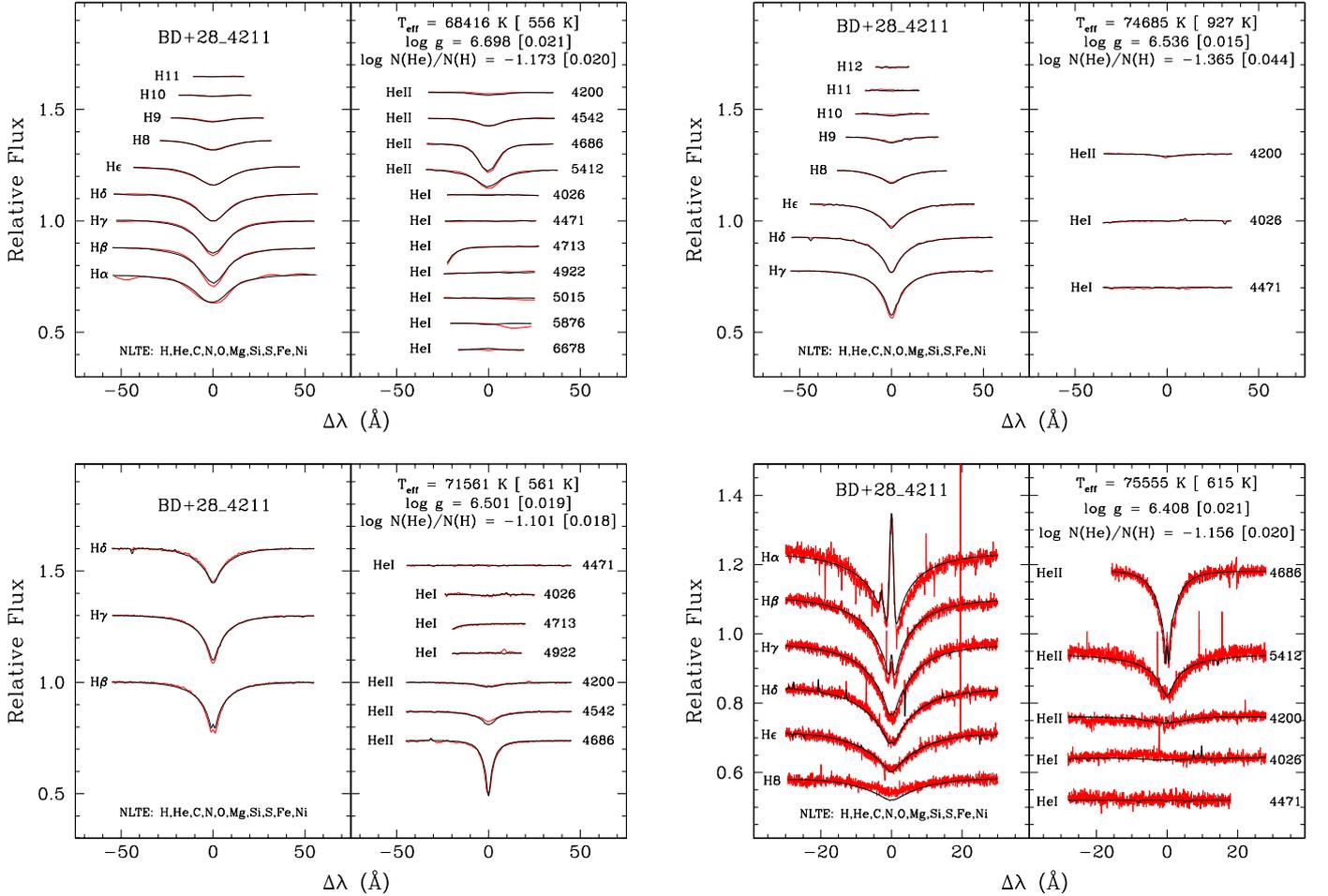

**Fig. 2.** Similar to Fig. 1, but using the NLTE model grid that includes the following elements, C, N, O, Mg, Si, S, Fe, and Ni, and the abundances determined in Paper I.

1992). Normalized lines of both the observed and model spectra (convolved at the instrumental resolution) are thus compared.

Resulting fits obtained with the metal-free grid are displayed in Fig. 1. Not surprisingly, the resulting temperature is much lower than what is expected from the UV analysis ($T_{eff}$ = 82,000 K ± 5000 K), but the other parameters are in acceptable ranges from the expected ones (log $g$ = 6.2$^{+0.3}_{-0.1}$, log $N$(He)/$N$(H) ~ −1.0). The resulting fits are rather bad and assessing parameters on such results is not a good option. Note, however, that the very high S/N of our three SO spectra helps a lot in terms of defining "badness" here. The fits of the BOK8.7 and BOK1.3 spectra show good examples of the so-called Balmer line problem, with the lowest lines in the series being too shallow in the model, while the trend shifts in H8 with a model line that is too deep. There is also a hint in the BOK8.7 spectral fit that the resulting temperature is too low when one looks at the helium lines: weak neutral helium lines are predicted by the model while the observed spectrum is merely flat at these wavelengths and, in addition, the two main ionized helium lines are not strong enough in the model spectrum. These shortcomings are also evident in the fit of the high resolution UVES spectrum, as well as large differences in the cores of several lines.

It should be pointed out that the uncertainties on the derived atmospheric parameters quoted in Fig. 1 (and following) are only the formal errors of the fit in 3D space. They do not take into account external errors and systematic effects. Ignoring differences in S/N, spectral coverage, and resolution from one spectrum to another, a better way of verifying the internal consistency of these results is to compute the mean value and the standard deviation for each of the parameter. For the metal-free grid, we thus obtain the following mean values (based on the 4 different spectral fits), $T_{eff}$ = 66,250 K ± 1053 K, log $g$ = 6.367 ± 0.081, and log $N$(He)/$N$(H) = −1.210 ± 0.080. This is reported, as well as the inferred parameters of the individual fits, in the top third of Table 1. Note that one obvious systematic effect that decreases somewhat the mean value of log $N$(He)/$N$(H) is related to the fact that the BOK1.3 spectrum covers only one He II line, and that is the weak 4200 Å feature.

The fits performed with the second grid having abundances fixed to the ones determined for BD+28°4211 in Paper I were expected to give more satisfying results. As compared to the metal-free case, the temperatures obtained are indeed higher, while the inferred surface gravities and helium abundances change only slightly and remain within the expected ranges. (see Fig. 2). Nevertheless, in spite of having included abundances that were self-consistently determined, and for which the main UV spectral lines were very well reproduced, the fitting procedure does not give appropriate effective temperatures. The results are too cool by roughly 10,000 K. A close look at the resulting fits shows that they are significantly improved as compared to the metal-free case, but they are not perfect either. A remnant of the Balmer line problem is still visible in our best fits, and the He II line at 5412 Å cannot be reproduced properly in the BOK8.7 spectrum. In addition, the details of the line core emission in both Hβ and He II at 4686 Åare not well modeled in the UVES spectrum; the discrepancies suggest again too low an inferred temperature. the results of the fits with our second model grid are reported in the middle third of Table 1. The mean values are $T_{eff}$= 72,554 K ± 2813 K, log $g$ = 6.536 ± 0.105, and log $N$(He)/$N$(H) = −1.199 ± 0.100.

At this point two things must be kept in mind. The first one is that in this range of parameters, $T_{eff}$ and log $g$, the Balmer and helium lines are only weakly sensitive to a change of parameters. A variation of effective temperature slightly changes the depth in the very core of the lines while the wings remain essentially the same. The surface gravity has a higher impact on the wings but also influences the depth, but again not by a very large amount (this will be discussed in more details in the next section). This causes an intrinsic uncertainty associated with any parameter determination based on optical data. Secondly, a consequence of the first point is that using very high quality spectra (in terms of S/N) allows to see the small discrepancies between our best-fit models and the observations. With spectra having a more representative sensitivity, the fits from Fig. 2 might have looked acceptable and the parameters thus deduced would have been wrong. It should also be mentioned that forcing the temperature to a value of 82,000 K, while fitting the two other parameters, does not result in much better fits with this grid of models. Such a fit can be seen in Figure 4 of Latour et al. (2014a).

### 3.3. Exploring Metallicity Effects on Spectral Lines

In his important contribution on the Balmer line problem in a NLTE context, Werner (1996) expressed the hope that including more elements than C, N, and O (like he did in his experiments) would definitely solve the problem in stars such as the hot sdOs that he investigated, BD+28°4211 itself and the similar object LS V+46°21 (the central star of S216). The inclusion of C, N, and O in solar amounts helped a lot, but there were still remaining discrepancies between the observed and computed line profiles that most likely were due to additional missing opacity. As shown just above, the inclusion of C, N, O, Mg, Si, S, Fe, and Ni with the specific abundances derived from our UV analysis of BD+28°4211 in Paper I did improve considerably the spec-tral fits, but our effort felt short in the sense that there is ample room still for improvement and we significantly underestimate the effective temperature of our target. This is somewhat disappointing, and we must conclude that significant opacity is still missing in these models[9].

Prior to the work of Werner (1996), Bergeron et al. (1993) had shown that the Balmer line problem can be solved in the hot DAO white dwarf Feige 55 ($T_{eff}$~ 60,300 K, log $g$ ~ 7.25) if an abundance of Fe equal to 25 times its solar value with respect to H by number is used in the computations of the atmospheric structure (in the LTE approximation). Of course, this suprasolar value of the abundance of Fe was only used as a proxy for the overall metallic opacity in the atmosphere of Feige 55 and has nothing to do with the real abundance of that particular element. But the calculations of Bergeron et al. (1993) certainly indicated that there is generally quite a bit of "missing" opacity in model atmospheres of hot stars and that this can have significant influence on the modeling of the optical lines of H and He.

A similar problem was also found and discussed by O'Toole & Heber (2006) and Geier et al. (2007) concerning the difficulty of fitting simultaneously the H I, He I, and He II lines in the optical spectra of sdOB subdwarfs ($T_{eff}$ between ~ 30,000 K and ~ 40,000 K). Their proposed solution was to arbitrarily increase the metal abundances in LTE models to 10 times their solar values. In this way, consistent and acceptable spectral fits could be obtained. Likewise, and more recently, Gianninas et al. (2010) concluded that using a boosted metallicity consisting of 10 times the solar values of C, N, and O in NLTE models of hot DAO white dwarfs could be a practical approach to the analysis of optical spectra of such stars.

In this spirit, we decided to explore the effects of increasing the abundances of metallic elements in our model atmospheres. For this purpose, we built a coarse grid of dedicated models including eight effective temperatures between 22,000 and 90,000 K. The temperature of the models between 40,000 and 90,000 K varies by step of 10,000 K, and they have log $g$ = 6.0 and log $N$(He)/$N$(H) = −1.0, roughly representative of hot sdOs with normal helium content. The two coolest models, in order to be more representative of EHB stars, have the following parameters: $T_{eff}$ = 22,000 K with log $g$ = 5.4, and $T_{eff}$ =30,000 K with log $g$ = 5.6, and they both have log $N$(He)/$N$(H) = −2.0. For each of these eight sets of parameters, we built metal-free model atmospheres and models including the line blanketing of C, N, O and Fe at one, two, five, ten, and fifteen times their solar abundances. All of these models were computed in NLTE and the synthetic spectra include only hydrogen and helium lines to avoid the presence of strong and unrealistic metallic lines. In order to inspect the metallicity effects, we present in the four panels of Fig. 3 the line profiles of the five lowest members of the Balmer series (Hα to Hε), of four He I lines (4026, 4471, 4713, and 5876 Å), and of three He II lines (4542, 4686, 5412 Å). Line profiles are displayed for five metallicities: metal-free, one, five, ten, and fifteen times solar. We omitted from plotting the two times solar case because the line profiles are already quite crowded. We find Fig. 3 particularly instructive.

The most important result of Fig. 3 that we want to emphasize is the saturation effect that can clearly be observed in the plot: the line profiles no longer change with increasing metallicity beyond a certain level. This is particularly evident for the 80,000 K model, of central interest here, for which practically no distinction can be made in the profiles computed with a background metallicity of 5× (green), 10× (magenta), or 15× (blue) the solar abundances of C,N, O, and Fe. This suggests to us a practical recipe to compute some sort of saturated metallicity in model atmospheres, namely multipling by, say, a factor 10 the solar abundances of metals. This is nothing more than the suggestion already put forward by O'Toole & Heber (2006), Geier et al. (2007), and Gianninas et al. (2010), but with the added justification of the saturation effect.

**Table 1.** Results of our fitting procedures for BD+28°4211

| Spectrum | $T_{eff}$ (K) | log $g$ (cm s$^{-2}$) | log $N$(He)/$N$(H) (dex) |
|---|---|---|---|
| NLTE H,He model grid | | | |
| BOK8.7 | 64,584 ± 1,113 | 6.463 ± 0.053 | −1.177 ± 0.048 |
| BOK1.3 | 67,504 ± 690 | 6.387 ± 0.020 | −1.336 ± 0.046 |
| MMT | 66,479 ± 659 | 6.238 ± 0.023 | −1.208 ± 0.024 |
| UVES | 66,432 ± 602 | 6.380 ± 0.022 | −1.118 ± 0.027 |
| rms | 66,250 ± 1053 | 6.367 ± 0.081 | −1.210 ± 0.080 |
| NLTE line-blanketed grid with the metallic abundances of BD+28°4211 | | | |
| BOK8.7 | 68,416 ± 556 | 6.698 ± 0.021 | −1.173 ± 0.020 |
| BOK1.3 | 74,685 ± 927 | 6.536 ± 0.015 | −1.365 ± 0.044 |
| MMT | 71,561 ± 561 | 6.501 ± 0.019 | −1.101 ± 0.018 |
| UVES | 75,555 ± 615 | 6.408 ± 0.021 | −1.156 ± 0.020 |
| rms | 72,554 ± 2813 | 6.536 ± 0.105 | −1.199 ± 0.100 |
| NLTE line-blanketed grid with ten times solar abundances | | | |
| BOK8.7 | 79,694 ± 1,332 | 6.508 ± 0.045 | −1.157 ± 0.033 |
| BOK1.3 | 80,678 ± 1,174 | 6.536 ± 0.015 | −1.380 ± 0.045 |
| MMT | 82,738 ± 639 | 6.582 ± 0.016 | −1.050 ± 0.013 |
| UVES | 82,257 ± 660 | 6.450 ± 0.019 | −1.152 ± 0.017 |
| rms | 81,342 ± 1219 | 6.519 ± 0.048 | −1.185 ± 0.121 |

---

[9]  In this context, we reemphasize that our neglect of the contributions of F, P, and Ar at their derived UV abundances in our calculations cannot be at the origin of the problem.

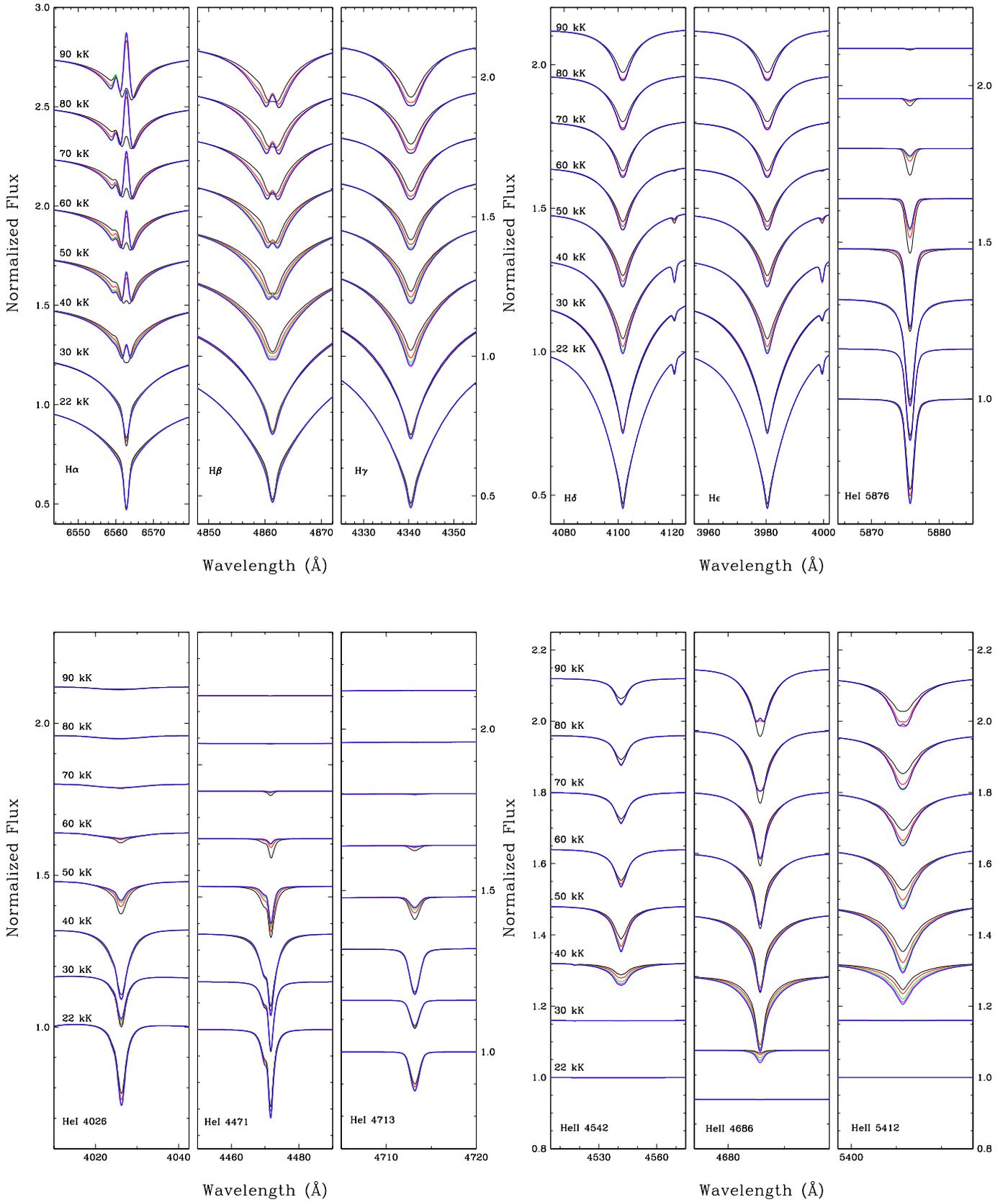

**Fig. 3.** Comparison of the line profiles for our models with different metallicities: metal-free (black), one (red), five (green), ten (magenta), and fifteen (blue) times the solar abundances for C, N, O, and Fe. Each line profile is normalized and the vertical scale is adjusted for each lines in order to allow the best view of the profiles. The synthetic spectra are convolved at a resolution of 1.0 Å.

Figure 3 further reveals that the saturation effect is verified over the full range of parameters displayed in the plot. The largest differences between the line profiles are found for the 40,000 K model, and, yet, the differences between the 10× (magenta) profiles and the 15× (blue) profiles in that particular model remain tiny for all of the lines illustrated. As for the sdB domain, sampled by our two coolest models (with different log $g$ and log $N$(He)/$N$(H) values than the hotter models), the line profiles are clearly not very sensitive to the assumed background metallicity. This result is not new in a sdB star context and has been rediscussed recently (see, e.g, Figs. 2 and 4 of Latour et al. 2014b).

### 3.4. Spectral Analysis with Metal-Enhanced Models

In the light of the results obtained in the previous subsection, we decided to test this concept of enhanced metallicity. We thus computed a third full grid of model atmospheres dedicated to the spectral fitting of BD+28°4211. It is similar to our second grid, except that the "minimal" metallicity of the latter (defined by the UV abundances of C, N, O, Mg, Si, S, Fe, and Ni) is replaced by a "saturated" metallicity (defined by ten times the solar abundances of these 8 elements). Technically speaking, it takes substantial amounts of time to build grids of NLTE models with nonzero metallicity, but since the metallicity has to be progressively "turned on" in our approach, the availability of the second grid helped us save considerable time in our passage from minimal to saturated opacity. Note also that in the computation of the synthetic spectra of this third grid (which is made with SYNSPEC), the metal abundances were reduced to the ones in BD+28°4211 to avoid unrealistic and strong metallic features in the optical spectra. In other words, the artificially enhanced metallicity was used only in the computation of the atmospheric structures (with TLUSTY).

It is instructive to compare the temperature stratifications of atmosphere models having the same values of the effective temperature, surface gravity, and helium content, but obtained from the three different grids. As an example, Fig. 4 illustrates the effects of metals on the temperature structure of model atmospheres having fundamental parameters representing BD+28°4211: $T_{eff}$ = 82,000 K, log $g$ = 6.4 and log $N$(He)/$N$(H)= −1.0. The first model is a metal-free one, showing the typical NLTE temperature inversion in the outer layers of the atmosphere (dotted line). Adding the metallic content of BD+28°4211 causes a drastic cooling of the outer layers while the deeper ones are heated (dashed curve). We showed in Paper I that the cooling is essentially due to the C, N, O, elements while both these elements and Fe heat the inner layers. When looking at the temperature stratification for a metal-enhanced model (solid curve), the most striking effect is the significant warming, again of the inner layers, that is prominent in the line-forming region (between a depth of −1.0 and −2.0). This line-forming region can be localized with the help of the $\tau_\nu$ = 2/3 curve, which indicates the depth (in column density) at which half the photons leave the atmosphere at a given wavelength.

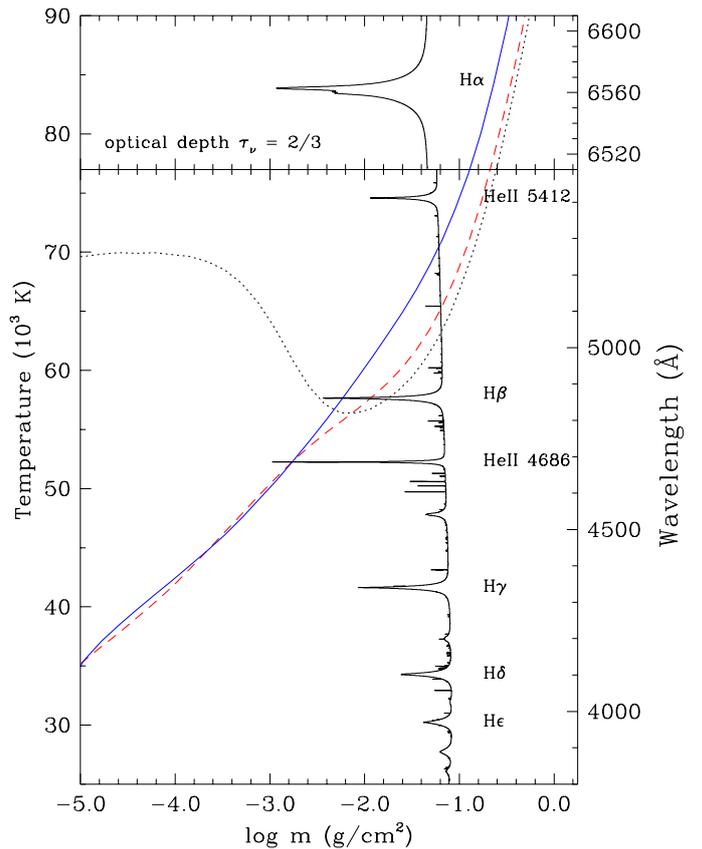

**Fig. 4.** Temperature stratification and monochromatic optical depth $\tau_\nu$ = 2/3 as functions of depth, where $m$ is the column density, for NLTE models defined by $T_{eff}$ = 82,000 K, log $g$ = 6.4, and log $N$(He)/$N$(H) = −1.0. The temperature structure is shown for three model atmospheres having different compositions: with H and He only (black, dotted), with the metallic abundances of BD+28°4211 (red, dashed) and with ten times solar abundances (blue, solid). The $\tau_\nu$ = 2/3 curve is from the latter model and shows wavelength intervals corresponding to the Balmer line series.The wavelength sampling of this curve is about 0.3 Å.

In the last step of the procedure, we fitted our four reference optical spectra using the third grid of models. The resulting fits can be seen in Fig. 5. These fits are rather remarkable in the sense that they reproduce very well *all* of the available observed line profiles, including details of line core emission in the high resolution UVES spectrum. Moreover, despite using spectra of different sensitivity, spectral coverage, and resolution, the inferred atmospheric parameters fall, in all four cases, within the ranges of values derived from the detailed UV analysis presented in Paper I. To our knowledge, this is the first time that realistic estimates of the atmospheric parameters of a hot sdO star, especially the effective temperature, have been obtained through the application of a simultaneous fit of all available H and He lines in a given optical spectrum, a method initially put forward by Bergeron et al. (1992) in the white dwarf context. With emphasis, we point out that this was possible only under the assumption of an increased metallicity.

The resulting parameters of our various fits are summarized in the lower third of Table 1. The mean values are $T_{eff}$ = 81,342 K ± 1219 K, log $g$ = 6.519 ± 0.048, and log $N$(He)/$N$(H) = −1.185 ± 0.121. Of particular interest, this new estimate of the effective temperature is well within the well-constrained value of $T_{eff}$ = 82,000 ± 5000 K derived from the UV spectrum in Paper

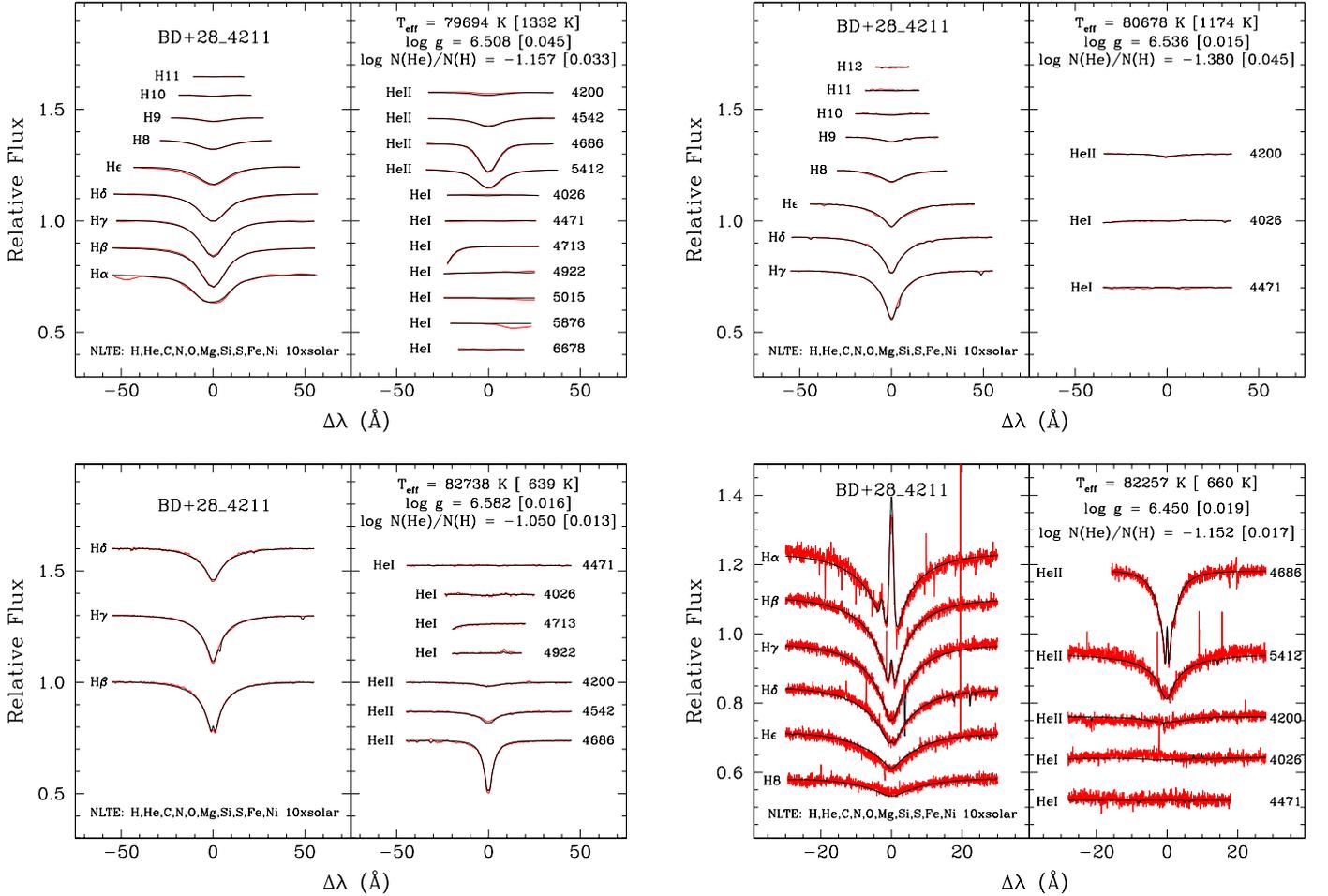

**Fig. 5.** Similar to Fig. 1, but using the NLTE model grid that includes the following elements, C, N, O, Mg, Si, S, Fe, and Ni, with solar abundances multiplied by a factor of 10.

I. For the surface gravity of BD+28°4211 a value of $\log g = 6.5$ is also compatible with the conclusions of Paper I which led to a formal estimate of $\log g = 6.2^{+0.3}_{-0.1}$, although the new spectroscopic value is just formally acceptable. In this context, let us remind the reader that the UV metallic lines (mainly iron ones) did not help in constraining very tightly the surface gravity, but a better match was nevertheless obtained with $\log g = 6.2$, hence the adopted value. On the other hand, when we compared the spectroscopic distance of BD+28°4211 for various combinations of masses and surface gravities with the one given by the Hipparcos parallax of the star, a $\log g \geq 6.4$ was needed, unless the mass of the star is significantly lower than the canonical value of ~0.5 $M_\odot$ for a hot subdwarf. This is why the upward uncertainty we adopted on $\log g$ allows for a surface gravity of at most 6.5. This upper value gave good spectroscopic distances for masses between 0.4 and 0.5 $M_\odot$ and still was not conflicting with the UV metal lines. We also checked the old Hipparcos parallax value for the star (Perryman et al. 1997) and the distance derived with this older value is between 88 and 126 pc. This is a little closer to the spectroscopic one but still does not allow for a good match of the distances with models having $\log g$ of 6.2, unless the mass is around 0.3 $M_\odot$. In the light of the present analysis, it is most likely that the true value of the surface gravity of BD+28°4211 is closer to 6.5 dex than to the value of 6.2 dex suggested in Paper I[10]. Finally, we note that a better estimate of the helium content is likely obtained if the BOK1.3 spectrum is excluded from the averaging process since the latter bears a very weak signature of the He abundance. One then gets $\log N(\text{He})/N(\text{H}) = -1.120 \pm 0.049$.

## 4. Additional Verifications

### 4.1. HIRES Spectra

It is interesting to compare selected line profiles for fiducial models of BD+28°4211 with those gathered from HIRES archives as discussed previously in Sect. 2. This is particularly true for the lines showing strong core emission at high resolution, namely He II 4686 Å, Hβ, and Hα. We have obtained a very nice global fit of the UVES data we modeled as shown in the lower right panel of Fig. 5, including those lines. However, the HIRES data allow further detailed comparisons because of their higher S/N (up to ~ 300) even though their resolution is slightly degraded (~ 0.1 Å) compared to the UVES. As discussed above, and contrary to the UVES data, the HIRES spectra are not suit-able for a formal multiline analysis, but we can still use them to test further our model atmospheres.

---

[10] Assuming $T_{\text{eff}}$ = 82,000 K, $\log g$ = 6.5, $M$ = 0.5 $M_\odot$ and the reddening discussed in Paper I, we find a spectroscopic distance to BD+28°4211 of $d = 112 \pm 40$ pc, which is indeed compatible with either the old, 88−126 pc, or new, 81−106 pc, Hipparcos distance.

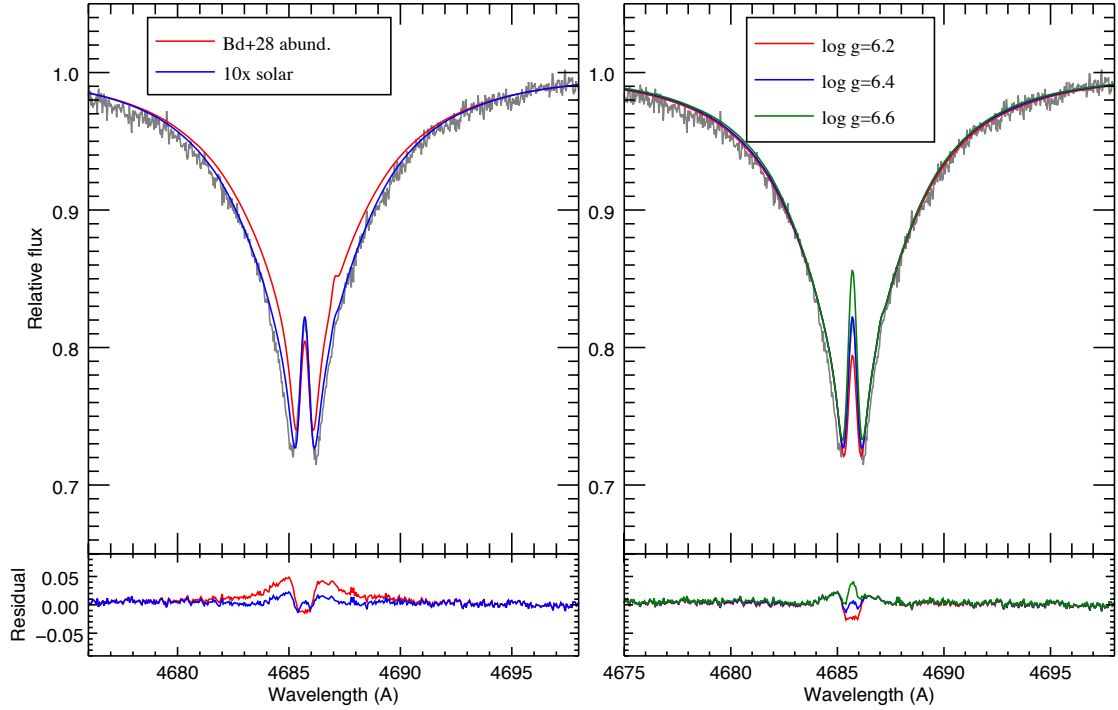

**Fig. 6.** Comparison between synthetic spectra and the He II 4686 Å line from HIRES (1997-08-12). *Left Panel*: Synthetic spectra from models at 82,000 K, $\log g = 6.4$, and $\log N(\mathrm{He})/N(\mathrm{H}) = -1.0$. In red, the spectrum comes from a model with the abundances of BD+28°4211 determined in Paper I. In blue, the spectrum is from a model with ten times the solar metallicity. *Right Panel*: Spectra from the ten times solar metallicity grid, having various surface gravities.

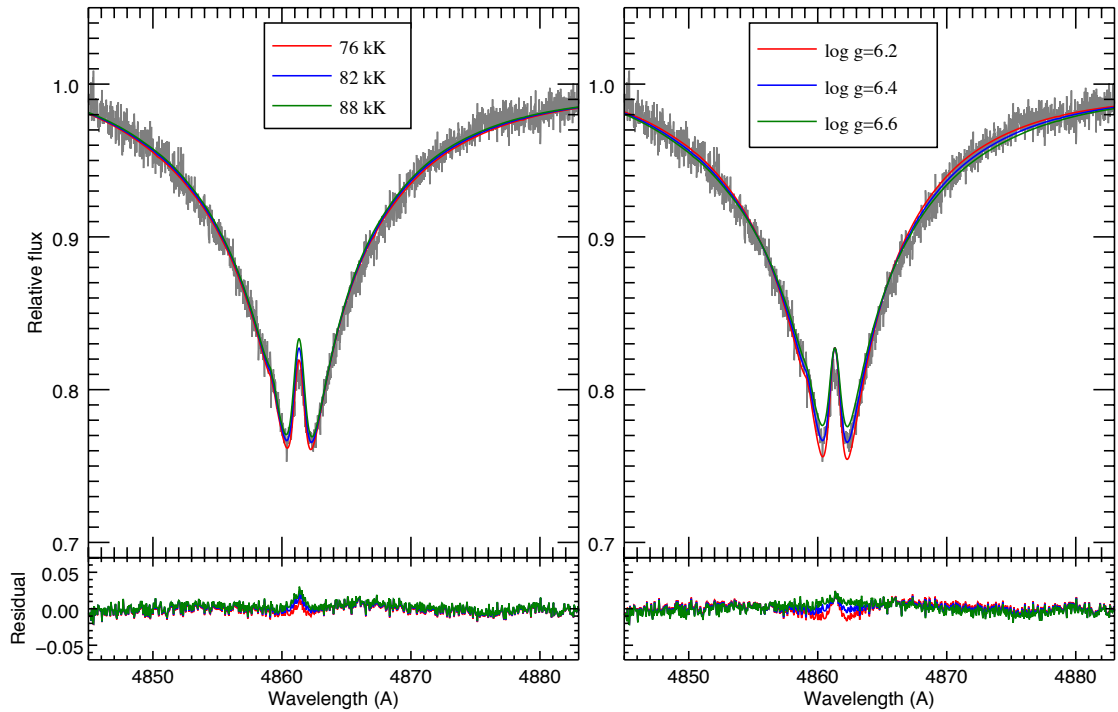

**Fig. 7.** Comparison between synthetic spectra and the H$\beta$ line from HIRES (2011-10-04). *Left Panel*: For models having different effective temperatures. *Right Panel*: For models having different surface gravities.

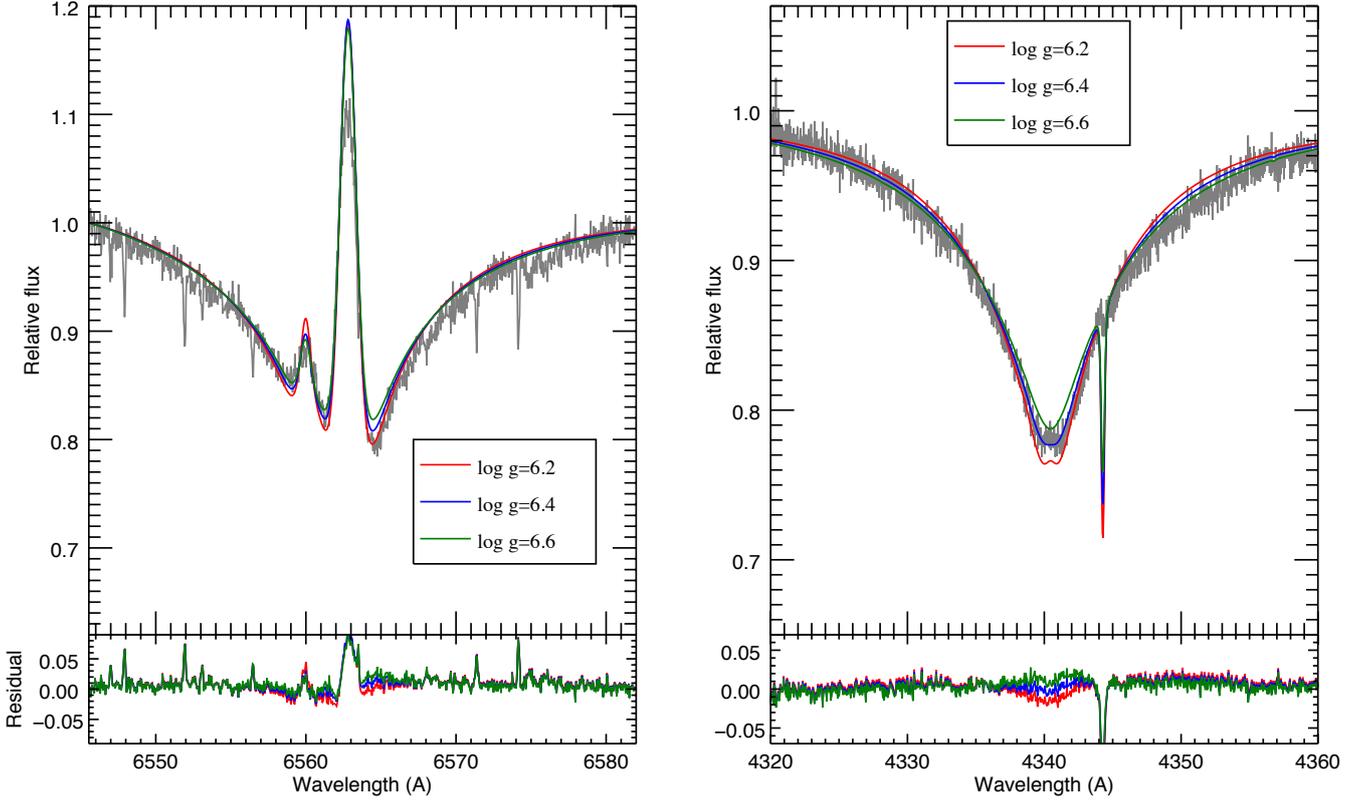

**Fig. 8.** *Left Panel*: Comparison between synthetic spectra having different log *g* values and the Hα line from HIRES (2011-10-04). *Right Panel*: Similar, but for the Hγ line.

Referring to the lower right panel of Fig. 3, the He II *λ*4686 line appears not much affected by a variation of the metallic content of the model atmospheres (from 1× to 15× the solar abundances), at least at the 1.0 Å resolution of these synthetic spectra. A closer look, with a HIRES spectrum boasting a tenfold increase in resolution, shows that there is a definite improvement in the way *λ*4686 is reproduced when the metallicity is increased. This is particularly well illustrated in the left panel of Fig. 6. Indeed, the match between a fiducial model spectra ($T_{\rm eff}$ =82,000 K, log *g* = 6.4 and log *N*(He)/*N*(H) = −1.0, metal abundances ten times solar) and the HIRES observation of this helium line is very good. In addition, we mentioned above that at such a high temperature, the strongest Balmer and helium lines are not very sensitive anymore to changes in log *g* or $T_{\rm eff}$. We explicitly illustrate this point in the right panel of Fig. 6 where model spectra having different values of log *g* between 6.2 and 6.6 are depicted. One can see that this particular line is rather insensitive to such a change, although the comparison favors the higher gravities.

We examined in a similar way Hβ in Fig. 7 where it is compared with models from the enhanced metallicity grid having different values of $T_{\rm eff}$ and log *g*. The left panel shows that Hβ is rather insensitive to changes of effective temperature while a change in the surface gravity has a small effect on the depth of the line (right panel). Note that in this comparison, as well as in the following ones in this subsection (unless otherwise stated), the values of the fixed parameters are among $T_{\rm eff}$ = 82,000 K, log *g* = 6.4, and log *N*(He)/*N*(H) =−1.0.

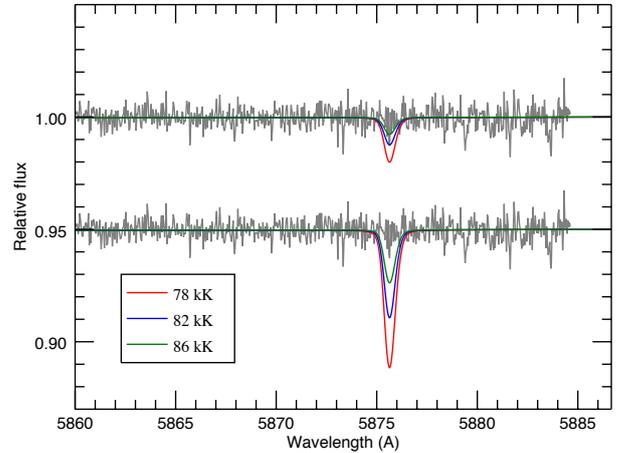

**Fig. 9.** Comparisons between synthetic spectra and the He I 5875 Å region from HIRES (1997-08-12) for models having different effective temperatures. Top – The model spectra are taken from our metal enriched grid. Bottom – The model spectra come from a metal-free model grid and have log *g* = 6.2 dex, as those used by Napiwotzki (1993).

The effects of varying the assumed value of log *g* on the line profiles of Hα and Hγ are illustrated in Fig. 8. As in the cases of He II *λ*4686 and Hβ just discussed, the effects are rather small, but the lower value of log *g* = 6.2 is disfavored. Note that the sharp absorption lines in the Hα region are telluric lines, so they do not originate from the star. Our comparison shows that the emission peak of our fiducial models with enhanced metallic-ity is too high compared to the data, but the depth and width of the line are well reproduced. This is also the case for the UVES fit illustrated in the lower right panel of Fig. 5, but is more difficult to see at the scale of that other plot. We also compared the line with models from our second grid, having the abundances of BD+28°4211, and the emission peak is also higher than the observed one, but the model lines are neither deep nor wide enough.

The comparison between Hγ and some of our models is shown in the right panel of Fig. 8. This line is also rather well reproduced with our models, but this time a surface gravity of 6.4 dex offers markedly a better match. The thing to note here is the presence of a tiny emission core in the observed line; such an emission is not often seen in Balmer lines other than Hα and Hβ. A hint for a very weak emission feature can also be detected in our UVES spectrum. The high resolution is essential to see this type of feature. Emission is barely seen in our lowest gravity model and it is not strong enough to reproduce the one observed.

There is one last feature that was worth investigating with the HIRES spectra and that is the He I 5875 Å line. Napiwotzki (1993) used this line to secure the effective temperature he deduced for BD+28°4211 using the Hε line. He compared model spectra having different temperatures with his observations and found a best match for an effective temperature between 80 and 85 kK (let us remember that metal-free NLTE models were used at the time). He had at his disposal a 0.4 Å resolution spectrum that appears to be more noisy than the HIRES ones. The feature he associated with He I λ5875 is barely visible in his spectrum but should be distinguishable in the HIRES ones. However, after a careful search we did not find any trace of this line after inspecting three different observations. If there is indeed a line, then it is weaker than the noise level. Figure 9 shows the region of interest for the observations of the 1997-08-12 night. The top comparison is with three models having different effective temperature, taken from our metal-enhanced grid. With $T_{\text{eff}} \geq$ 82,000 K, the line is predicted to be within the noise level. Since the helium line is not visible in the observations, we used the C IV emission line at 5811 Å (identified in Herbig 1999) present in the same order to accurately fix the wavelength scale in that spectral region. The bottom comparison is made with synthetic spectra analog to those used by Napiwotzki, i.e., metal-free and with a surface gravity of 6.2 dex and the lines predicted are way too strong.

The 23 spectra retrieved from the KOA span a 14 years time span, from August 1997 to October 2011, which is an interesting baseline to look for any long term radial velocity (RV) variations. The RVs were measured using between 8 and 20 metallic lines, depending on the wavelength coverage and S/N (see Herbig (1999) for a list of BD+28°4211 metallic lines). These individual RVs were then averaged out for every observation. The values obtained did not show any significant variations over these 14 years. The mean value of the RV for the 23 observations is 22.1 km s⁻¹ with a standard deviation of σ = 2.3 km s⁻¹, which is in agreement with those reported by Herbig (1999). It is thus very unlikely that BD+28°4211 is part of a binary system, unless the inclination is close to 0° or the orbital period is much longer than our baseline, in the latter case, evolutionarily speaking, it is essentially a single star.

### 4.2. UV Helium Lines

The previous comparisons between high resolution optical lines and our model spectra with ten times solar abundances clearly demonstrated that our metal-enhanced models, overall, match very well the Balmer and helium lines seen in the HIRES spectrum of BD+28°4211. But what about the helium lines in the UV range, are they significantly affected by an enhanced metallicity? This point is worth investigating because, referring to Figure 5 of Paper I in which an oxygen line is located just next the to He II 1640 Å, one can notice that the helium line in question seems quite well reproduced by the model used then, while we just saw that the optical He II lines cannot be reproduced with such models.

Figure 10 shows comparisons of this line, first with models having various log $g$ and the abundances of BD+28°4211, where one can see that the changes thus induced are rather small. The wings are well reproduced by the models but the central absorption is wider in the observations. This is a bit intriguing and we verified that no change of parameters ($T_{\text{eff}}$, $N(\text{He})/N(\text{H})$) affects much the width of the core. The rotational velocity of the star is known to be quite small (Herbig 1999), so this option can be disregarded. As for a metallicity effect, the right panel of Fig. 10 shows that there is only a slight difference in the line profile between a model having the abundances of BD+28°4211 and one having ten times the solar metallicity and this difference is not in the central core. We do not know why the central core is not reproduced correctly; it might have something to do with the theoretical line profile, since the core width remains unaffected by changes in the parameters of model atmospheres. We did not find in the literature any mention of this problem, but it would at least require the availability of a star having fundamental parameters relatively close to those of BD+28°4211 in order to display a similar line profile with a sharp central core absorption. Alternatively, microturbulence (not included in our synthetic spectra) might be at work here.

We finally also examined how the He II 1085 Å line featured in the FUSE spectrum of BD+28°4211 is reproduced by our model spectra. We compared it with models having the metallic abundances determined in Paper I and various surface gravities in Fig. 11. Again the variation of log $g$ does not produce important differences in the line profile and the models having realistic abundances reproduce well this helium line. This is true also for metal-enhanced models.

## 5. Discussion

We began our comprehensive analysis of BD+28°4211 with the ultimate goal of retrieving the atmospheric parameters of the star by carrying out a standard fitting procedure of its optical spectrum alone, as it is done for the cooler hot subdwarfs. The challenge was to somehow "overcome" the Balmer line problem that prevents from modeling the observed Balmer series with a unique set of fundamental parameters. As discussed in Sect. 1, this problem is a major one in hot sdO stars and makes it very tricky to determine fundamental parameters of such stars, especially their effective temperatures. A careful inclusion of metal line-blanketing in model atmospheres seemed a promising way to solve this issue, or at least to diminish the discrepancies between observed and theoretical lines as was shown by Werner (1996).

The first part of this analysis, presented in Paper I, focussed on the UV spectrum, which is a standard way for studying very hot stars and usually leads to sound results (Rauch et al. 2007; Ziegler et al. 2012; Rauch et al. 2013). By self-consistently fitting the numerous metallic lines in the UV spectra from FUSE and the HST spectrograph STIS, we were able to draw up a realistic chemical composition for the atmosphere of BD+28°4211.

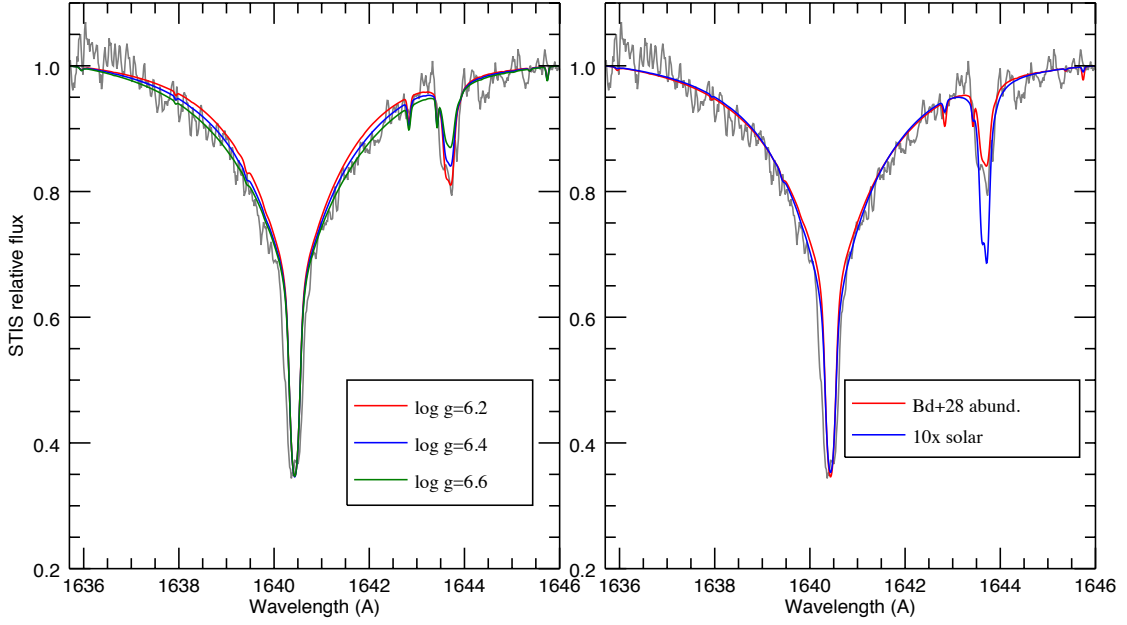

**Fig. 10.** Comparisons between synthetic spectra and the He II 1640 Å line from the STIS spectrum. *Left Panel*: With models having metallic abundances corresponding to those of BD+28°4211, $T_{eff}$ = 82,000 K, log $N$(He)/$N$(H) = −1.0 and various log $g$. *Right Panel*: With models having log $g$ = 6.4 but different metallic contents.

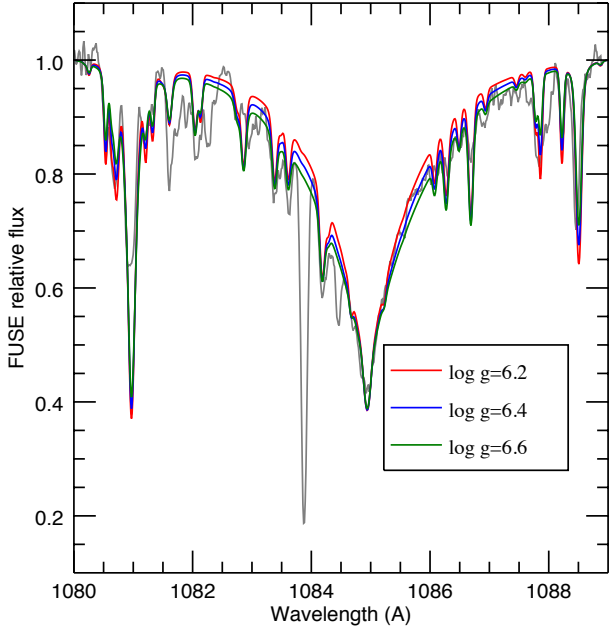

**Fig. 11.** Comparisons between synthetic spectra and the He II 1085 Å line from the FUSE spectrum. The models have the metallic content of BD+28°4211 and different values of log $g$.

We also confirmed that the previously estimated fundamental parameters ($T_{eff} \sim$ 82,000 K, log $g \sim$ 6.2, solar $N$(He)/$N$(H)) are in good agreement with the observed UV spectrum. The following step in our study, the subject of this article, was to use the abundances thus determined to build a grid of NLTE line-blanketed model atmospheres specifically suited to BD+28°4211 and use it to perform a spectroscopic fit of the star's optical spectrum. However, even though significant improvements were obtained in the modeling of the observed optical lines as compared to the case where metal line-blanketing was neglected, we disappointedly realized that these custom-made models were still falling short of the desired results. And indeed, as seen in Fig. 2 for example, the best fits obtained with our four spectra give effective temperatures too low by about 10,000 K and, in addition, the observed spectral lines are not well reproduced in their details. However, note in this latter case that with much lower S/N data than our superlative spectra (BOK8.7, BOK1.3, MMT), it would have been very difficult, if not downright impossible, to detect the small but quite significant differences between the observed and computed line profiles in the panels of that figure.

In the past, some similar "fitting" problems were solved by artificially increasing the metallicity of LTE model atmospheres. This method proved useful in the sdOB transition range ($T_{eff} \sim$ 35,000 K) where the LTE metal-rich models (with ten times the solar metallicity) yield improved matches between the observed lines and the best-fitting models, but this without changing much the fundamental parameters thus derived (O'Toole & Heber 2006; Geier et al. 2007). In that temperature range, the mismatch between observed and model spectra is not so much in the Balmer lines themselves but rather in the helium ones, for which the lines originating from both ionization stages cannot be correctly reproduced without this artefact. The Balmer line problem is also seen in hot white dwarfs and a few of those analyzed in Gianninas et al. (2010) were much better reproduced with NLTE model atmospheres including C,N,O with 10 times their solar abundances. It is thought that this approach compensates, at least partially, for some unknown opacity sources to the point where the atmospheric structure is affected in the "correct" way.

With these informations in mind, we next investigated the effects of varying the background metallicity on the optical lines of interest. In this context, Fig. 3 is an important result of our present work. The plot indeed shows that the line profiles saturate beyond a certain value of the assumed metallicity. In particular, there are practically no differences in line profiles obtained in a wide range of $T_{\mathrm{eff}}$ for a metallicity defined by 5×, 10×, or 15× the solar abundances of C, N, O, and Fe. This suggests that the concept of saturated metallicity could be used as an interim cure for the missing opacity problem in hot subdwarf stars. In practical terms, full grids of model atmospheres should be computed with the help of this artefact in order to analyze various samples of optical spectra, the latter ideally being characterized by a high sensitivity and/or high resolution.

As for BD+28°4211, we thus built a dedicated grid of NLTE line-blanketed model atmospheres including the 8 most abundant metals found in Paper I, but having ten times their solar abundances to make sure the saturated regime was reached. This metal-enhanced grid ultimately allowed us to derive satisfactory fundamental parameters for BD+28°4211 on the basis of our optical spectra. Our best estimates, based on the straight average of the results derived from four optical spectra of different sensitivity, spectral range, and resolution, give $T_{\mathrm{eff}} = 81,342 \, \mathrm{K} \pm 1219 \, \mathrm{K}$, $\log g = 6.519 \pm 0.048$, and $\log N(\mathrm{He})/N(\mathrm{H}) = -1.120 \pm 0.049$, with the uncertainty being the standard deviation of the results. These are perfectly compatible with the results of Paper I ($T_{\mathrm{eff}} = 82,000 \pm 5,000 \, \mathrm{K}$, $\log g = 6.2^{+0.3}_{-0.1}$, $\log N(\mathrm{He})/N(\mathrm{H}) = -1.0$ [assumed]), which is based on the standard UV approach for very hot stars. The higher suggested surface gravity now solves the apparent conflict discussed in Paper I between the spectroscopic and Hipparcos distances. Specific tests indicated that the most important spectral features "pulling" towards a higher surface gravity are H$\delta$, H$\epsilon$, and He II at 4542 Å.

As a posteriori test, we exploited some of the HIRES data of BD+28°4211 available in the KOA. With their high resolution and good S/N, they provide an incomparable insight into the detailed profiles of several Balmer and helium lines of the star. We were able to overcome the drawback that comes with these data, the wavy continuum, and ended up with observed lines that could be compared with our models. That way we tested if our optimal model atmospheres, with their artificially enhanced metal abundances, could reproduce the detailed observations of HIRES. Our comparisons showed that our metal-enriched models indeed reproduce very well the following lines : He II $\lambda 4686$, H$\alpha$, H$\beta$, and H$\gamma$. In the details, there is a small discrepancy between our models and the emission peak observed in H$\alpha$, which is predicted to be higher. A tiny emission bump is also discernible in the core of H$\gamma$ but was not fully reproduced by our models. The radial velocities measured for a sample of HIRES spectra, covering a 14 years time period does not show any significant variation within 5 km s$^{-1}$, thus indicating that the star is most likely a single one.

What we learned from our analysis is that despite the fact that the UV spectrum can be very well reproduced by model atmospheres including the metal abundances derived for BD+28°4211, such models fail to reproduce the optical Balmer and helium lines. In order to achieve adequate results in the optical domain, we had to include in our models an artificially enhanced metallicity. The good side to this is that we were then able to derive appropriate fundamental parameters for BD+28°4211, based only on its optical spectrum. The downside is that to get these results we had to set our metal abundances to unrealistic values. These large abundances somehow affect the atmospheric structure of our models in a way that makes the optical lines correctly modeled. It is likely that the additional blanketing brought about by the enhanced abundances of the species included in our models account for some missing opacities present in the star but not in our models. Our use of what we called a saturated metallicity should only be seen as a proxy for some important missing physics. It is possible that these missing opacities come from atomic species not included in our models (these species should not be dominant in the star numberwise, but their opacities might be important), transition lines not accounted for, or improper broadening of some metal lines. Likewise, it is possible that incorrect opacity sampling might be at work here. With the presence of spectral lines originating from trans-iron elements, such as Ge, Ga, As, Sn, and Pb, in the spectra of hot subdwarf stars as well as in those of a few hot white dwarfs (O'Toole 2004; Naslim et al. 2011; Werner et al. 2012; Rauch et al. 2015 and references therein), it is possible that the opacity of such elements constitute a part of the missing opacity. Reindl & Werner (2015) also showed that Ne in solar abundances can lead to an important change in the temperature structure of the atmosphere in a 100 kK model. However, it seems a little odd that such missing opacities would significantly affect the Balmer and helium optical lines, while the UV ones can be accounted for very well without the induced change in the atmospheric structure. In any case, this knowledge should be very useful for obtaining more accurate fundamental parameters for hot sdO stars (and possibly also for very hot white dwarfs) when observations in the optical range are the only ones available.

In the light of our results and of previous investigations, we thus propose, along with earlier researchers, that the atmospheric structures of hot stars be computed with artificially enriched metal abundances as an interim solution for estimating their atmospheric parameters if only optical spectroscopy is available. We propose our concept of saturated metallicity for the whole domain of hot subdwarf stars. In particular, the procedure should be applied to the case of the newly discovered pulsating stars in $\omega$ Cen (Randall et al. 2011) which are among the rare sdO stars known to pulsate. Their temperature determination via the fitting of their Balmer and helium lines with a grid of NLTE line-blanketed model spectra with normal metallic abundances yield values around 50,000 K. However, preliminary non-adiabatic exploration of the sdO star region did not show pulsational instabilities around this particular effective temperature (Randall et al. 2012), but only at higher values. A legitimate question that might be raised in this case is about the validity of the spectroscopically derived temperature, which must certainly be underestimated according to the present findings. We hope that the upcoming UV observations of two $\omega$ Cen pulsators will settle the issue and allow us to test our approach with these stars, but we have to keep in mind that their optical spectra is of limited quality given the relative faintness of the stars. The sdO star Feige 34 would also be a good candidate to test our approach, like BD+28°4211 it is a spectroscopic standard for which good observational data (UV and optical) are available. Its optical spectrum suggests that it is cooler than BD+28°4211, but still quite hot ($T_{\mathrm{eff}} \sim 70,000$ K).

Finally, we summarize our main results as follows.

We fitted high-quality spectra of BD+28°4211 using NLTE model atmospheres including the metallicity determined in Paper I via our UV analysis.

The best fits obtained with these models were improved when compared to fits made with models only including H and He, but they indicate a temperature too low by 10,000 K, and do not perfectly reproduce the spectral lines.

We investigated the effect of increasing our model atmospheres metallicity (up to 15 times solar) on spectral lines for a wide range of temperatures. We observed for most of them a saturation effect at ten times solar metallicity; beyond this value, the line profiles no longer change.

We adopted this ten times solar metallicity to build a new metal-enhanced grid. The fitting procedure then led to very good fits and accurate atmospheric parameters.

We then compared our new best-fit models with high-resolution, high S/N observed spectra culled from archived HIRES observations and found very good agreement.

We therefore propose metal-enriched model atmospheres (ten times solar) for determining the fundamental parameters of hot stars when only optical spectroscopy is available. This should lead to more realistic parameters than using models with normal metallic content.

*Acknowledgements.* This work was supported in part by the NSERC Canada through a fellowship awarded to M.L. and through a research grant awarded to G.F. The latter also acknowledges the contribution of the Canada Research Chair Program. M.L. also acknowledges funding by the Deutsches Zentrum für Luft- und Raumfahrt (grant 50 OR 1315). We thank L. Fröhling for sharing her RV measurements and P. Németh for interesting discussions. This work has made use of the Keck Observatory Archive (KOA), which is operated by the W. M. Keck Observatory and the NASA Exoplanet Science Institute (NExScI), under contract with the National Aeronautics and Space Administration. This paper also used data obtained from the ESO Science Archive Facility. We are also grateful to the PIs of the HIRES and UVES observations we used.

# References

Bergeron, P., Saffer, R. A., & Liebert, J. 1992, ApJ, 394, 228
Bergeron, P., Wesemael, F., Beauchamp, A., et al. 1994, ApJ, 432, 305
Bergeron, P., Wesemael, F., Lamontagne, R., & Chayer, P. 1993, ApJ, 407, L85
Dreizler, S. & Werner, K. 1993, A&A, 278, 199
Edelmann, H., Heber, U., Hagen, H.-J., et al. 2003, A&A, 400, 939
Fontaine, G., Green, E. M., Brassard, P., Latour, M., & Chayer, P. 2014, in Astronomical Society of the Pacific Conference Series, Vol. 481, 6th Meeting on Hot Subdwarf Stars and Related Objects, ed. V. Van Grootel, E. M. Green, G. Fontaine, & S. Charpinet, 83
Fontaine, M., Chayer, P., Oliveira, C. M., Wesemael, F., & Fontaine, G. 2008, ApJ, 678, 394
Geier, S. 2013, A&A, 549, A110
Geier, S. & Heber, U. 2012, A&A, 543, A149
Geier, S., Heber, U., Podsiadlowski, P., et al. 2010, A&A, 519, A25
Geier, S., Nesslinger, S., Heber, U., et al. 2007, A&A, 464, 299
Gianninas, A., Bergeron, P., Dupuis, J., & Ruiz, M. T. 2010, ApJ, 720, 581
Haas, S., Dreizler, S., Heber, U., Jeffery, S., & Werner, K. 1996, A&A, 311, 669
Heber, U. 2009, ARA&A, 47, 211
Heber, U., Edelmann, H., Lisker, T., & Napiwotzki, R. 2003, A&A, 411, L477
Herbig, G. H. 1999, PASP, 111, 1144
Hirsch, H. 2009, PhD thesis, Friedrich-Alexander-Universität Erlangen-Nürnberg
Hubeny, I. & Lanz, T. 1995, ApJ, 439, 875
La Palombara, N., Esposito, P., Mereghetti, S., & Tiengo, A. 2014, A&A, 566, A4
Lanz, T. & Hubeny, I. 1995, ApJ, 439, 905
Latour, M., Fontaine, G., Brassard, P., et al. 2011, ApJ, 733, 100
Latour, M., Fontaine, G., Chayer, P., & Brassard, P. 2013, ApJ, 773, 84
Latour, M., Fontaine, G., Chayer, P., Brassard, P., & Green, E. 2014a, in Astronomical Society of the Pacific Conference Series, Vol. 481, 6th Meeting on Hot Subdwarf Stars and Related Objects, ed. V. van Grootel, E. Green, G. Fontaine, & S. Charpinet, 67
Latour, M., Fontaine, G., Green, E. M., Brassard, P., & Chayer, P. 2014b, ApJ, 788, 65
Latour, M., Randall, S. K., Fontaine, G., et al. 2014c, ApJ, 795, 106
Lemke, M. 1997, A&AS, 122, 285
Massey, P. & Gronwall, C. 1990, ApJ, 358, 344
Moehler, S., Dreizler, S., Lanz, T., et al. 2011, A&A, 526, A136
Napiwotzki, R. 1992, in Lecture Notes in Physics, Berlin Springer Verlag, Vol. 401, The Atmospheres of Early-Type Stars, ed. U. Heber & C. S. Jeffery, 310–+

Napiwotzki, R. 1993, Acta Astronomica, 43, 343
Napiwotzki, R. & Rauch, T. 1994, A&A, 285, 603
Naslim, N., Jeffery, C. S., Behara, N. T., & Hibbert, A. 2011, MNRAS, 412, 363
O'Toole, S. J. 2004, A&A, 423, L25
O'Toole, S. J. & Heber, U. 2006, A&A, 452, 579
Perryman, M. A. C., Lindegren, L., Kovalevsky, J., et al. 1997, A&A, 323, L49
Ramspeck, M., Haas, S., Napiwotzki, R., et al. 2003, in Astronomical Society of the Pacific Conference Series, Vol. 288, Stellar Atmosphere Modeling, ed. I. Hubeny, D. Mihalas, & K. Werner, 161
Randall, S. K., Calamida, A., Fontaine, G., Bono, G., & Brassard, P. 2011, ApJ, 737, L27+
Randall, S. K., Fontaine, G., Calamida, A., et al. 2012, in Astronomical Society of the Pacific Conference Series, Vol. 452, Fifth Meeting on Hot Subdwarf Stars and Related Objects, ed. D. Kilkenny, C. S. Jeffery, & C. Koen, 241
Rauch, T., Rudkowski, A., Kampka, D., et al. 2014, A&A, 566, A3
Rauch, T., Werner, K., Bohlin, R., & Kruk, J. W. 2013, A&A, 560, A106
Rauch, T., Werner, K., Quinet, P., & Kruk, J. W. 2015, ArXiv e-prints [[arXiv]1501.07751]
Rauch, T., Ziegler, M., Werner, K., et al. 2007, A&A, 470, 317
Reindl, N. & Werner, K. 2015, in 19th European Workshop on White Dwarfs, ed. P. Dufour, P. Bergeron, & G. Fontaine, in Press
Saffer, R. A., Bergeron, P., Koester, D., & Liebert, J. 1994, ApJ, 432, 351
Stroeer, A., Heber, U., Lisker, T., et al. 2007, A&A, 462, 269
Tremblay, P.-E. & Bergeron, P. 2009, ApJ, 696, 1755
van Leeuwen, F. 2007, A&A, 474, 653
Werner, K. 1996, ApJ, 457, L39
Werner, K., Rauch, T., Ringat, E., & Kruk, J. W. 2012, ApJ, 753, L7
Ziegler, M., Rauch, T., Werner, K., Köppen, J., & Kruk, J. W. 2012, A&A, 548, A109